\documentclass[acmsmall]{acmart}

\AtBeginDocument{%
  \providecommand\BibTeX{gin{%
    \normalfont B\kern-0.5em{\scshape i\kern-0.25em b}\kern-0.8em\TeX}}}

\usepackage{CJKutf8}

\begin{document}

\setcopyright{acmlicensed}
\acmPrice{15.00}
\acmDOI{10.1145/3611059}
\acmYear{2023}
\copyrightyear{2023}
\acmSubmissionID{play23main-p6878-p}
\acmJournal{PACMHCI}
\acmVolume{7}
\acmNumber{CHI PLAY}
\acmArticle{413}
\acmMonth{11}
\received{2023-02-21}
\received[accepted]{2023-07-07}

\title[Understanding the Practices and Experience of Danmaku Participation Game Players in China]{Let's Play Together through Channels: Understanding the Practices and Experience of Danmaku Participation Game Players in China}

\author{PiaoHong Wang}
\email{piaohwang2-c@my.cityu.edu.hk}
\affiliation{%
  \institution{City University of Hong Kong}
  \city{Hong Kong}
  \country{China}
}

\author{Zhicong Lu}
\affiliation{%
  \institution{City University of Hong Kong}
  \city{Hong Kong}
  \country{China}
}
\email{zhiconlu@cityu.edu.hk}

\begin{abstract}

Live streaming is becoming increasingly popular in recent years, as most channels prioritize the delivery of engaging content to their viewers. Among various live streaming channels, \textit{Danmaku participation game} (DPG) has emerged in China as a mixture of live streaming and online gaming, offering an immersive gaming experience to players. Although prior research has explored \textit{audience participation games} (APGs) in North America and Europe,
it primarily focuses on discussing prototypes and lacks observation of players in natural settings. Little is known about how players perceive DPGs and their player experience. To fill the research gap, we observed a series of DPG channels and conducted an interview-based study to gain insights into the practices and experiences of DPG players. Our work reveals that DPGs can effectively synergize live streaming and online games, amplifying both player engagement and a profound sense of accomplishment to players. 


\end{abstract}



\received{21 February 2023}
\received[revised]{2 June 2023}
\received[accepted]{7 July 2023}


\maketitle

\section{Introduction}

In recent years, video game live streaming has been developing rapidly as a popular form of entertainment. Twitch, the most successful video game live streaming platform, had 6.9 million monthly active streamers and 2.1 million concurrent viewers in 2020 \cite{wolff2022audience}. Among various video game channels on Twitch, the \textit{Audience Participation Game} (APG) channels enable viewers to participate in video games through channels instead of merely being passive audiences. These streaming channels could collect viewers' real-time text comments and execute the instructions represented by these comments inside the game \cite{striner2021mapping}. These APG channels adopt specific controller algorithms to synthesize viewers' comments from channels to make the final decisions in video games \cite{lessel2017crowdchess}. This novel design made APG channels attracted massive attention in 2014. As one of the earliest APGs, ``Twitch Plays Pokémon'' enabled viewers to play \emph {Pokémon Red} in the live streaming channel together~\cite{ramirez2014twitch}, and it successfully broke the Guinness World Record for having ``the most participants on a single-player online video game''. However, APGs developed slowly in recent years after a short period of fast development. Compared to well-developed Twitch categories (e.g. ``League of Legends'') that have tens of millions of followers, there are only four hundred thousand APG followers and fewer than 30 active APG channels on Twitch in 2022.

\vspace{2mm}
\begin{figure} 
\centering 
\includegraphics[width=10cm, height=4.9cm]{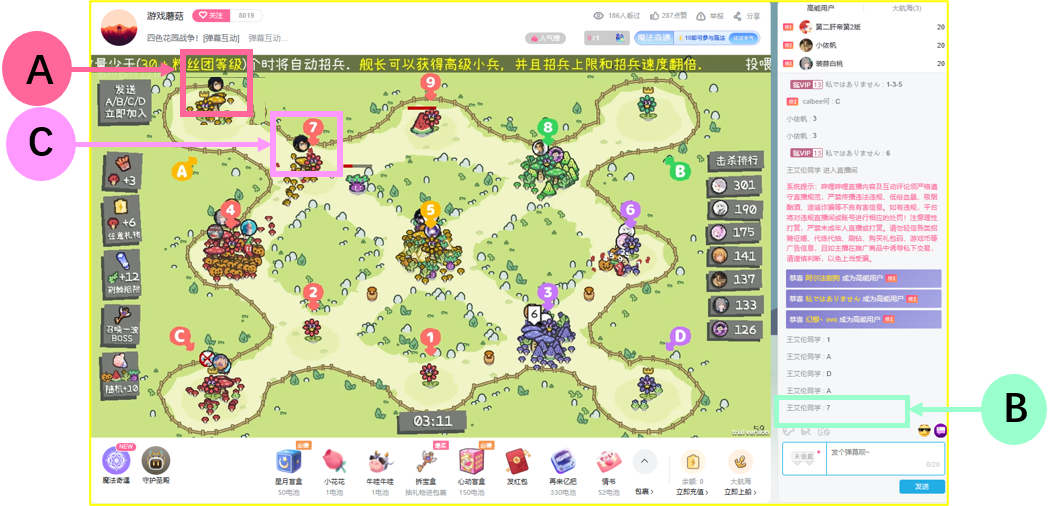} 
\caption{The \emph{Garden Fight} channel on Bilibili. A)The independent game character assigned to the player. B)The player inputs a comment in the live streaming channel to command the assigned character to move to the number 7 crossroad. C)The character moved to the crossroad after receiving the new command}
\label{introduction} 
\end{figure}

In February 2022, Danmaku Participation Games (DPG) gradually gained popularity on Bilibili \footnote{https://www.bilibili.com/}, a video-sharing and live streaming platform in China. The platform is famous for Danmakus, textual comments that appear overlaid on top of videos or live streams on East Asian video-sharing platforms \cite{wu2018danmaku}. Unlike most APGs aggregating comments from players, DPG is one kind of special emerging APGs on Bilibili which allows players to play games through controlling independent characters in live streaming channels. For example, \emph {Puppy Disco}, one of the most popular DPG channels allowing players to control their characters to dance in the channel, has attracted more than 800K subscribers. Without the controllers for aggregating comments in APG channels, DPGs are more similar to massive multiplayer online role-playing games (MMORPGs), allowing players to interact freely with other players in live streaming channels. Most related studies in HCI and game research primarily focused on mainstream crowd-based APGs. However, DPGs which could afford individual independent control were under-studied. To understand why DPGs are popular in China as well as the distinctions between DPGs and mainstream APGs with collective control and to inform the CHI Play community about the unique player practices and experiences brought by these distinctions, our study aims to answer the following research questions: 

\begin{itemize}

     \item \textbf{RQ1:} What are the main categories of DPGs, and what are typical mechanisms of DPGs?
     
     \item \textbf{RQ2:} What unique player interactions and experiences can be supported by DPG channels? 
     
\end{itemize}

To address the above questions, we observed 50 DPG channels and conducted an interview-based study with 13 DPG players. In this paper, we first review the literature on APGs, followed by the introduction of the playing process of DPGs on Bilibili (e.g., the steps for selecting and joining the DPG channels). Then we report the special player interactions and experiences afforded by DPGs (e.g., a strong sense of engagement). Our findings show that DPGs are distinct from APGs with aggregative control in several attributes, mainly the identities assigned to players, the interaction among players afforded by channels, and the relationship between streamers and channel players. On the one hand, these distinctions allow players to interact freely with each other and enjoy a higher level of engagement compared to crowd-based APGs. On the other hand, these innovations also force novice players to accept the supervision of skilled ones and endure the influence on their gaming experience from other players in the channels. Our work also suggests opportunities to provide a complete gaming experience to players who can hardly afford high-end devices or live with busy schedules. Thus, our contributions to HCI are i) a detailed description of DPG channels on Bilibili and ii) a nuanced understanding of the experiences and practices of DPG players. Our work fills an important gap in HCI by reconsidering how live streaming could be better integrated with online games for a better gaming experience.

\section{RELATED WORK}

We first review the research into video game live streaming and audience participation games to anchor the key contributions of our work based on prior research.

\subsection{Video Game Live Streaming and Beyond}

Live streaming has existed for over two decades as a form of real-time entertainment, which provides equal opportunities for everyone to present themselves online \cite{smith2013live}. Since the outbreak of COVID-19 in 2020, some offline on-site activities, such as folk clubs   \cite{benford2021producing} and fitness classes \cite{guo2022s}, have also been increasingly transformed to online via live streaming. Live streaming platforms (e.g., Twitch) have attracted countless streamers so anyone can find suitable channels and live content they want \cite{wolff2022audience}. Because of the unique affordances, streaming has gained traction in the HCI community in recent years \cite{tang2017perspectives,faklaris2016legal,lu2018you,scheibe2016information}. Not just for fun and entertainment, users may engage with live streaming for educational \cite{chen2021towards,hammad2021towards,streamwiki2018lu}, creativity\cite{streamsketch2021lu}, cultural \cite{lu2018you,ICH2019lu,livestreaming2020lu, vtuber2021lu}, commercial~\cite{cai2018utilitarian,ecommerce2022tang,ecommerce2023wu}, or other purposes. Among various types of channels, video game live streaming is increasingly gaining traction, which
enables audiences to enjoy watching the gameplay of live streamers   \cite{smith2013live}. The live streaming websites allow viewers to interact with streamers by inputting text chat messages \cite{wulf2020watching,kaytoue2012watch}. Enhanced channels providing better viewer-streamer interaction during the gaming process are also proposed \cite{lessel2017expanding}.
Researchers also explored factors that affect the experience of viewers \cite{lessel2018viewers}, how online communities form based on live streaming \cite{hamilton2014streaming}, and the design of more advanced live streaming systems to preserve performative information and accommodate participatory roles for different stakeholders \cite{lytle2020toward, Yen2023Storychat}. With the development of technology, the scope of live streaming games is also expanding. In addition to regular video games, VR games have gained traction in recent years \cite{emmerich2021streaming}. Although many researchers have studied how to increase viewer's sense of engagement in live streaming channels, the viewers in above literature still have limited impact on the gaming process. Our work firstly reveals that viewers could also act as independent players to enjoy themselves in live streaming channels by uncovering a unique genre of live streaming, i.e., Danmaku Participation Game, focusing on the unique features of such channels and both the players' practices and experiences with this new live streaming genre.

\subsection{Audience Participation Game (APG) and Danmaku Participation Game (DPG)}


Prior research argues that bidirectional interactions between live streaming and viewers can help create authentic social connections \cite{striner2021mapping}. Audience Participation Games (APGs) allow viewers to interact with the game rather than only act as audiences. The viewers may determine the challenges for streamers \cite{Legend, Choice} or control the character together \cite{Pokemon,ramirez2014twitch}. For this type of crowd-base APGs, researchers have discussed APG design space \cite{striner2021mapping}, development toolkit \cite{glickman2018design}, potential design challenges \cite{seering2017audience}, and the characteristics of audience participation \cite{cerratto2014understanding}. 
As a particular type of interactive game, APG also allows streamers to understand their viewers \cite{9231771}. In previous studies, APGs could be seen as special live streaming channels allowing players to play specific games together by interacting with streamers or shared game controllers. Previous studies have identified five main attributes of most APGs \cite{striner2021mapping,seering2017audience,Pokemon}: i) Support for channel game playing, ii) Control through inputting comments, iii) Identity assigned to channel players, iv) Dynamics between streamers and players, and v) Game-related interactions afforded by channels. These attributes distinguish APG channels from regular live streaming channels, which could only offer passive entertainment. To promote a better player experience, researchers have introduced AI technologies to facilitate interactions in APGs \cite{paliyawan2022audience,paliyawan2020towards}. Since most APGs are crowd-based, the controllers are necessary for aggregating commands from viewers and making final decisions \cite{lessel2017crowdchess}. The controller mechanism makes the player and channel connection extremely fragile. The participation will break down when the number of viewers who input commands in the channel is too large \cite{striner2021mapping}. With many simultaneous input commands, it is difficult for individual players to feel their influence on the games. This weakness undoubtedly limits the development of APGs. As of 2022, the APG category has only 400K followers on Twitch. In addition, the shared controllers force the players to share one gaming identity and make it hard for them to interact with each other in the channels. On the contrary, Danmaku Participation Games (DPG) appeared on Bilibili as one type of special APGs in 2022, allowing viewers to participate in games independently by inputting Danmaku in the chat box. Each viewer can play a role (e.g., a cartoon character, a spaceship, or a puppy) to interact freely with the game and other viewers within the live streaming channel. Because of this advantage, DPG achieves a great success. For instance, \emph {Puppy Disco}, one of the most popular DPG channels on Bilibili, has attracted more than 800K subscribers. 

According to the APG design space proposed by Stringer et al \cite{striner2021mapping}, there are 19 themes grouped by three sensitizing concepts in APG design. However, many of the themes in their design space are not completely applicable to DPGs. The first difference concerns the involvement of streamers in APG and DPG channels. According to Stringer's design space, streamers are always significant parts of APGs. They should interact with players actively in the channels and please them during the gameplay. However, most of DPG channels do not require the involvement of streamers. Instead, players in DPG channels interact with each other, enjoy themselves, and maintain the order of channels together. The owners of the live streaming channels often just need to ensure that the games are executed properly on their devices rather than appearing in the channels to play with viewers as regular streamers. This feature largely allows DPGs to reduce their reliance on streamers. The second difference concerns the identity of players during the gaming process. In APGs, players are considered as a whole as a participant. Therefore, the APG designers always need to consider how to arrange for players to work together to push the game forward, such as letting players vote in the channels to determine the next step. In contrast, players in DPG channels are always assigned independent identities during the gaming process. The previous studies on APGs always focus on the dynamics between players and streamers as well as better aggregating input of players in the channels, which do not apply to DPG players since they can enjoy themselves instead of relying on others. It is still unknown how players experience, perceive, and practice DPG in the wild. Our study firstly fills this gap and uncovers the unique player experience and practice that DPG could afford. In addition, we found that DPGs could successfully address the challenges presented by prior researchers in APG. This work contributes to the broader game and play research in HCI by understanding how live streaming could be better integrated with online games and provide players with a complete and realistic gaming experience.

\section {METHODOLOGY}

To get a better understanding of DPGs, we first joined and observed 50 DPG channels on Bilibili. Following previous research on APGs \cite{glickman2018design}, we then conducted interviews with 13 DPG players to gain a deeper understanding of their gaming experience and practices. Same as Morva's work on Pokémon Go \cite{saaty2022pokemon}, we also adopted a thematic analysis approach \cite{braun2012thematic} to analyze the notes from observation and playing process as well as the transcripts of interviews with DPG players.

\subsection{DPG Channel Collection and Observation}

To create a comprehensive related list of DPG channels, we adopted the PRISMA systematic review procedure \cite{moher2009preferred}. Our first step was to identify potential DPG channels on Bilibili. For the purpose of our work, we only focus on game live streaming channels which allow viewers to participate in the games. Our systematic search was conducted on Bilibili and the external channel searching engine \emph{JustLive\footnote{http://live.yj1211.work/index/home/recommend}}. Because the search algorithm in Bilibili only shows channels that are live streaming, we relied on \emph{JustLive} to discover and record channels that are not streaming when we search and record the channels. We used the following keywords related to DPG: ``Danmaku Interaction'', ``Interaction'', ``Danmaku'', ``Interaction Game'', ``Audience Participation'', ``Danmaku Participation''. As our search progresses, we keep enriching the keyword list by adding the ones found from new channels. Then, we add all the search results to the related channel list. Next, we expand our list by adding ones recommended by Bilibili and JustLive. We analyzed two rounds of searching for all the channels found via the keywords or until all the recommended channels have been already added to the list. In the next step, we excluded the channels that were off during our observation process because we could not participate in these channels to observe them before we finalize the channel selection process. Finally, we excluded the channels that did not allow individual viewers to control specific characters. These channels were excluded after we identified their rules through participating in the channel game. This ensures that all the channels we reviewed are DPG channels. 

After identifying the DPG channels, one of our authors joined each channel and played with other online players for three days continuously and for over an hour each time, similar to the previous study in the MMORPG study \cite{ducheneaut2004social}. To guarantee the findings' reliability, all gaming processes were carried out on one computer with a stable Internet connection. The observing and playing process lasts three weeks, during which we take notes based on our observations and playing experience. Our observation focused on the channels' content (e.g., game content) and other factors that may distinguish DPG from APG (e.g., interaction among players \cite{zhuang2007player}). For example, we recorded the player chat messages and the interesting mechanisms of channels in our notes. All the notes were combined with interview transcripts together for further analysis.

\subsection{Qualitive Interviews with DPG Players}

\subsubsection{Interviewee Recruitment} To further understand the practice and experience of DPG players, we conducted a qualitative interview-based study with 13 DPG players. Since DPG has been a successful category on Bilibili, we believe it is appropriate to recruit DPG players from Bilibili through direct messaging. Our message contains a brief introduction of our project as well as our research team. We also describe in the message how much time the interviews will likely take up and the reward we could offer them. Three interviewees were recruited in this way. We also joined the communities established by each DPG channel owner and recruited another five interviewees from their communities. The recruitment advertisements were posted in Chinese on the community bulletin boards after we got permission from community administrators. The first author also posted the advertisement on other related social media platforms (e.g., \emph{DouBan Game Community}, one popular online forum for discussing anything related to games) and recruited five interviewees.

\begin{table}[]
\caption{Summary of participants interviewed.} 
\scalebox{0.80}{
\begin{tabular}{c|c|c|c|c}
\hline
ID                        & Age                       & Sex                      & Education level                   & Favourite Channels                            \\ \hline
1                         & 23                        & F                        & Bachelor's                        & \emph{Qingming Shanghe Tu}                              \\
2                         & 22                        & M                        & College student                   & Unrevealed                         \\
3                         & 24                        & F                        & Bachelor's                        & \emph{Mine Clearance}                                \\
4                         & 23                        & F                        & Bachelor's                        & \emph{Starcraft Fight}                              \\
5                         & 15                        & F                        & High school student               & \emph{Tank Fight}                               \\
6                         & 25                        & F                        & Bachelor's                        & \emph{Dash to the Underground}    \\
7                         & 22                        & M                        & College student                   & \emph{Were Wolf}                  \\
8                         & 18                        & F                        & College student                   & \emph{Army Fight}                              \\
9                         & 15                        & M                        & Middle school student             & \emph{Commander War}                             \\
10                        & 29                        & M                        & Bachelor's                        & Unrevealed \\
11                        & 23                        & M                        & Bachelor's                       & \emph{Zombie War}   \\  12                        & 23                        & F                        & Bachelor's               & \emph{Were Wolf} \\
13                        & 27                        & M                        & Master's               & \emph{Sakura}
\\ \hline
\end{tabular}}
\end{table}

\subsubsection{Interviewee Protocol} We conducted semi-structured interviews with 13 DPG players. The interviews were conducted by phone or text chat from May 2022 to February 2023. Each interview lasted from 20 minutes to 1 hour. All the interviewees get paid 15-100 yuan as their reward. The interviews include questions about their motivations to participate in DPG, which DPG channels they like best and why, how they think DPG differs from APG and ordinary MMORPG, and so on. Interviewees were also asked to describe the social interactions they were involved in DPG channels and their unsatisfying gaming experiences in DPG channels. When our interviewees compared DPGs to other games (e.g., APGs) during the interview process, we asked follow-up questions about how these distinctions might influence their gaming experience. The transcripts were recorded by the author in real time and confirmed with interviewees through real time communication to ensure that the general meaning of the transcribed text was correct, which greatly avoids the confusion caused by accents and noise. Some of the interviewees made supplements through text chat or phone call after the interviews (e.g., most are supplements concerning specific cases), which are also recorded by the author and added to the corresponding section of previously transcribed documents. All final transcribed materials were used for further analysis.  

\subsection{Data Analysis}

Two authors adopted the thematic analysis approach to analyze the notes and transcripts \cite{saaty2022pokemon}. The notes and transcripts were combined together during our analysis process. For example, when the interviewees mentioned their interactions with other players, we might compare their statements with the notes we obtained from our observation concerning the communication among players in channels. We start with open coding in two phases \cite{xiao2020random}. In the first phase, we coded the collected notes and transcripts line-by-line to closely reflect our data. Examples of such codes include “communicate with developers” and “fight with malicious players”. In the second phase, we focused on synthesizing the codes in the first phase and extracting high-level themes. Examples of these high-level categories include “interactions”, “experiences” , and “gaming process”. Our coding process was iterative; we continuously reviewed our data and codes to explore patterns in our data and synthesize our coding results to higher themes. All materials of our research were qualitatively coded by two researchers. During this process, two researchers discussed to resolve their disagreements on coding until reaching a consensus.

\vspace{2mm}
\begin{figure}[t] 
\centering 
\includegraphics[width=9cm, height=5cm]{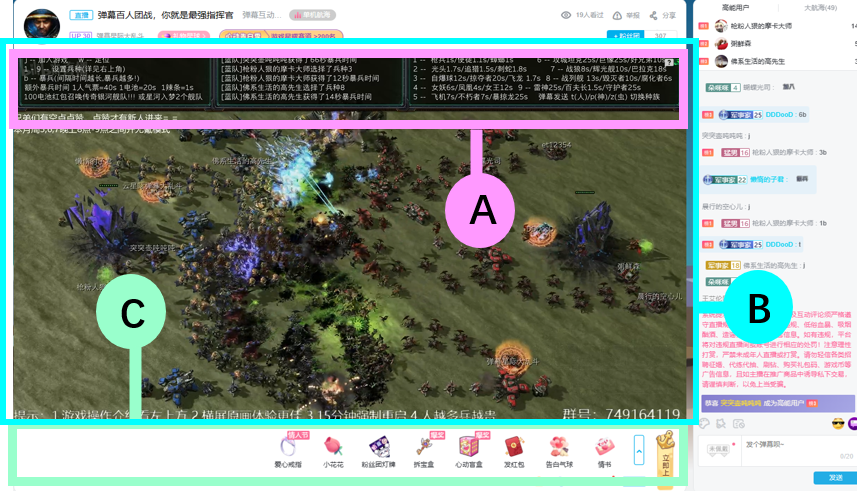} 
\caption{The interface of DPG channels on Bilibili: A). The guidance and tips from channel owners. B). \emph{Star Craft} channel live streaming video for displaying the game. C). The purchasable tools menu allows players to enhance their gaming experience by buying tools.} 
\label{Interface}
\end{figure}

\section{FINDINGS}

We now first elaborate on the process of playing DPGs on Bilibili based on our observation. We then describe the unique interactions among players and the player experience which could be afforded by DPGs based on both the observation and interviews.

\begin{figure}[t] 
\centering 
\includegraphics[width=14cm, height=5cm]{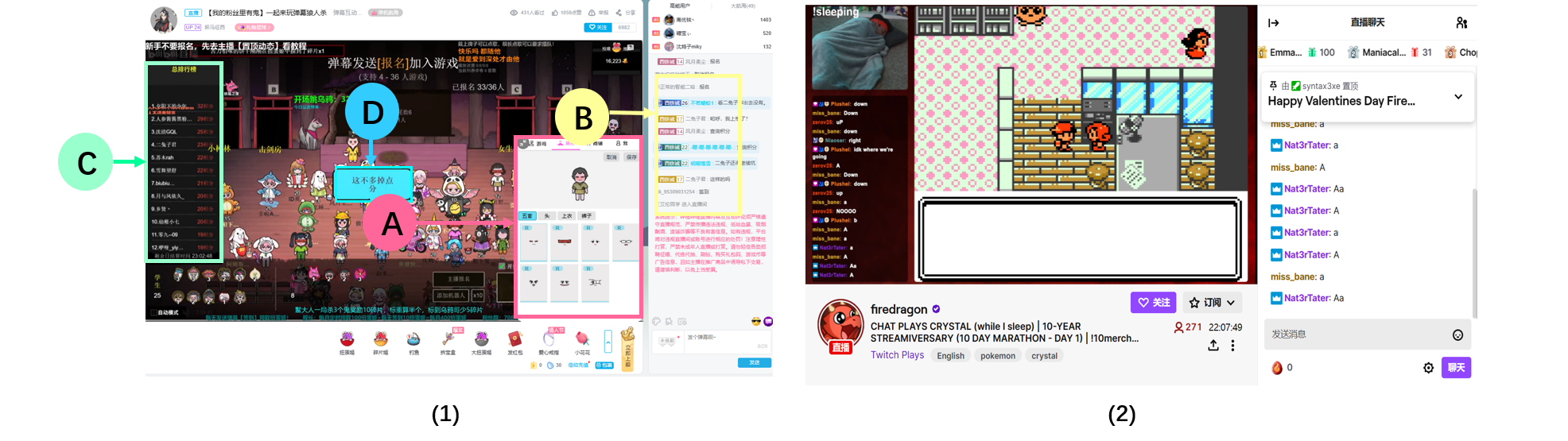} 
\caption{A comparison between APG and DPG in the player interface: 1). \emph{Were wolf}, a popular DPG on Bilibili. A). The players' mouse can click the character costume design interface. B). Honorary badges for players with high achievement. C). Leader board for all players in the channel. D). Public in-game communication dialogue in the channel (``Our points are going down'' in Chinese.), which is the chat among players in the channels. 2). \emph{Twitch Pokémon}, a typical APG on Twitch. Players in this channel must control the character together to drive the game forward (e.g., Defeating other characters' Pokémon), which differs from DPG.}
\label{Pokémon}
\end{figure}

\begin{table}[]
\caption{Summary of different categories of DPG content} 
\scalebox{0.85}{
\begin{tabular}{|c|c|l|}
\hline
Genre of Channel Content          & Representatives & \multicolumn{1}{c|}{Overview}                                                                                                                                                                                                                                                              \\ \hline
\emph{Real-Time Strategy DPG (RTS-DPG)} & \emph{Soldier Fight}   & \begin{tabular}[c]{@{}l@{}}Most RTS channels divide players into different teams \\ and forces them to compete with each other. The play-\\ ers need to input different commands (e.g.,``A13'' means \\ training 13 soldiers) for controlling their units (e.g., ca-\\ mp) to battle enemies.\end{tabular} \\ \hline
\emph{Shooting Game DPG (STG-DPG)}     & \emph{Tank Fight}      & \begin{tabular}[c]{@{}l@{}}Each player will be assgined a shooter (e.g., tank) and \\ they need to control the shooter for attacking other pl-\\ ayers through inputting commands (e.g.,``T-N13'' means \\ shooting at an angle of 13 degrees).\end{tabular}                                              \\ \hline
\emph{Fighting Game DPG (FTG-DPG)}     & \emph{Naruto}          & \begin{tabular}[c]{@{}l@{}}Each player controls a character and fight with other \\ players through inputting different action commands \\ (e.g., punching, kicking and jumping).\end{tabular}                                                                                                           \\ \hline
\emph{Puzzle Game DPG (PZL-DPG)}       & \emph{Sudoku Master}   & \begin{tabular}[c]{@{}l@{}}Each player will be assigned a place (e.g., a chessboard) \\ to play the puzzle game independently through input-\\ ting commands (e.g.,``A12C32D3'' means fill 3 in the 1o-\\ cation ``A12C13'').\end{tabular}                                                                   \\ \hline
\emph{Card Game DPG (CAG-DPG)}         & \emph{Wizard Fight}    & \begin{tabular}[c]{@{}l@{}}Each player acts as card master and controls what cards \\ are played each round for destroying other players'ca-\\ rds (e.g., ``A'' for selecting cards)\end{tabular}                                                                                                         \\ \hline
\emph{Music Game DPG (MSC-DPG)}        & \emph{Girl Disco}      & \begin{tabular}[c]{@{}l@{}}Players can control their characters' dance poses and \\ background music through inputting commands (e.g.,\\ ``Move to center of the stage'')\end{tabular}                                                                                                                     \\ \hline
\emph{Role-playing DPG (RPG-DPG)}      & \emph{Were Wolf}       & \begin{tabular}[c]{@{}l@{}}Players need to take on different roles (e.g.,wolf, doctor \\ and civilian) and achieve their goals through discussing \\ and reasoning (e.g., wolf need to kill all civilians, and \\ civilians need to discover all wolfs before all of them b-\\ eing killed)\end{tabular} \\ \hline
\end{tabular}}
\label{Games}
\end{table}

\subsection{Playing Process and Main Categories of DPG on Bilibili}

In this section, we first conclude the playing process and main categories of DPGs. Then we introduce how DPGs are distinct from APGs with collaborative control concerning the design of significant attributes.

\subsubsection{Elements of DPG Channel User Interfaces}

We follow the same framework in previous work on MMORPG interface study \cite{hsu2009exploring} to conclude the key elements of DPG channels. As depicted in \autoref{Interface}, there are five main components of the DPG channel interfaces: 

\begin{enumerate}
    \item Chat/Message/Comment inputting entry: On the right side of the channel, players can input comments to control their character and send messages to others in the channels (\autoref{introduction}.B, players could input command in the chat box to control 
    their characters or chat with other players in the channels).
    \item  Operation guidance: In the top left corner of the channel, it clearly shows what kind of comments players can enter to command their characters to do (For example, players inside the channel in \autoref{Interface} could see the guidance in \autoref{Interface}.B.).
    \item Purchasable tools menu: At the bottom of the channels, which offers a range of items that allow players to purchase to enhance their in-game character (\autoref{Interface}.C, players could buy tools in the bottom tool menu.). 
    \item Bulletin board: Anywhere on the central screen, which is always tips from channel owners/game developers
    (\autoref{Interface}, there are some warnings at the bottom of central screen.). 
    \item Character information and status: The information is always placed at the top of each player's character. Unlike regular APG, each player is assigned an independent character identified by each player's username to control (Figure 1.A). 
\end{enumerate}

\noindent Although DPGs miss many of the MMORPG interface elements (e.g., external/system interface \cite{hsu2009exploring}), which still allows players to perform basic social activities (e.g., grouping and chatting) and game operations (e.g., creating new characters), on the other hand, the DPG channel owners also attempt to dilute the limitations of the DPG interface through some design tricks (e.g., guidance in the channels to compensate for the lack of novice instructional content).

\subsubsection{DPG Genres on Bilibili}

As depicted in \autoref{Interface}, the players must first enter the channels they prefer on the Bilibili website. Consistent with previous studies about video game classification \cite{arsenault2009video}, we also adopt the same strategy to classify different DPG channels based on their game content. We found that DPGs can present diverse genres of games to players, which means the players can enjoy a completely different gaming experience from different channels. For example, in the STG-DPG channels, players could control the shooting tanks by inputting comments for adjusting the missile launch angle and strength. On the other hand, in RPG-DPG channels, players need to adjust their strategies by inputting additional comments for controlling the character to defeat specific enemies. In general, we identified seven different genres of DPG: (I). Real-Time Strategy DPG (RTS-DPG), (II). Shooting Game DPG (STG-DPG), (III). Fighting Game DPG (FTG-DPG), (IV). Puzzle Game DPG (PZL-DPG), (V). Card Game DPG (CAG-DPG), (VI). Music Game DPG (MSC-DPG) and (VII). Role-playing DPG (RPG-DPG). The overview description of these DPG channels is listed in \autoref{Games}.  

\vspace{2mm}
\begin{table}[t]
\caption{Comparison between APGs with collaborative control and DPGs in our study concerning the design of key attributes.}
\scalebox{0.75}{
\begin{tabular}{|c|l|l|}
\hline
{\color[HTML]{000000} APG Features}                           & \multicolumn{1}{c|}{{\color[HTML]{000000} APGs with collaborative control}}                                                                                          & \multicolumn{1}{c|}{{\color[HTML]{000000} DPGs in our study}}                                                                                                    \\ \hline
{\color[HTML]{000000} Support for channel game playing}       & {\color[HTML]{000000} \begin{tabular}[c]{@{}l@{}}The live streaming channels nee-\\ d to support players to be abl-\\ e to participate in games \cite{glickman2018design,striner2021mapping} \end{tabular}} & {\color[HTML]{000000} \begin{tabular}[c]{@{}l@{}}The live streaming channels co-\\ uld support players to play ga-\\ mes as independent players\end{tabular}}        \\ \hline
{\color[HTML]{000000} Control through inputting comments}     & {\color[HTML]{000000} \begin{tabular}[c]{@{}l@{}}The players control the games t-\\ ogether through voters \cite{ramirez2014twitch,lessel2017crowdchess} \end{tabular}}                                      & {\color[HTML]{000000} \begin{tabular}[c]{@{}l@{}}The players in the channels can \\ control their characters indep-\\ endently\end{tabular}}                     \\ \hline
{\color[HTML]{000000} Identity assigned to channel players}   & {\color[HTML]{000000} \begin{tabular}[c]{@{}l@{}}Players in the channels share o-\\ ne identity during the gaming p-\\ rocess in the channels \cite{ramirez2014twitch} \end{tabular}}                   & {\color[HTML]{000000} \begin{tabular}[c]{@{}l@{}}Each player in the channels is \\ assigned independent identity\end{tabular}}                                   \\ \hline
{\color[HTML]{000000} Dynamics between streamers and players} & {\color[HTML]{000000} \begin{tabular}[c]{@{}l@{}}Streamers always need to show \\ up in the channels and ammuse\\ players in the channels \cite{glickman2018design,striner2021mapping} \end{tabular}}       & {\color[HTML]{000000} \begin{tabular}[c]{@{}l@{}}Much lower reliance on streame-\\ rs. The players in the channels\\ can enjoy themselves.\end{tabular}} \\ \hline
{\color[HTML]{000000} Interaction afforded by channels}       & {\color[HTML]{000000} \begin{tabular}[c]{@{}l@{}}Interaction mainly happen be-\\ tween streamers and players \cite{striner2021mapping} \end{tabular}}                                    & {\color[HTML]{000000} \begin{tabular}[c]{@{}l@{}}Interaction can happen between\\ any players \end{tabular}}                                                      \\ \hline
\end{tabular}}
\label{comparison}
\end{table}

\subsubsection{Onboarding and Playing in DPG Channels}

After entering the DPG channel, the players must join the game by inputting specific comments which represent participation in the games. Otherwise, they can only be spectators. After successfully joining the game, players need to read the guidance prepared by developers in the channel, which explains what actions each comment will cause the character to perform. Most commands towards the character controlling are encoded into alphabetic and number strings (e.g., "12B" controls the character to fight with enemy ``B''). If the players enter incorrect comments, which means the comments cannot match any string commands for character controlling, the system may ignore their input. For example, the players in \emph{Garden Fight} channel could be assigned independent characters (Figure 1A) after joining the channel game. Then the players could control their characters to move to different spots (Figure 1C) through inputting real-time comments in the live streaming channels (Figure 1B).

\subsubsection{The Distinctiveness Between DPGs and APGs with Collaborative Control}

By comparing the observation notes we collected with previous APG studies, we found significant distinctions between DPGs and APGs with collaborative control in some key design attributes. Based on previous work concerning APGs with collaborative control \cite{glickman2018design,striner2021mapping}, we select five main attributes which could reflect the main differences between DPGs and APGs with collaborative control. Some of the attributes are adapted from more than one attributes in previous work to better reflect and summarises the significant differences between DPGs and APGs with collaborative control (e.g., we combine ``Relationship design between the streamer and the participant'' in \cite{glickman2018design} and ``Streamer/Viewer Relationship'' in \cite{striner2021mapping} into ``Dynamics between streamers and players'' in our work since there's an overlap between them.). As depicted in \autoref{comparison}, DPGs not only inherit the main attributes of APGs (e.g., support for channel gaming) but also innovate in the specific design of these attributes, distinguishing them from crowd-base APGs in previous studies. In contrast to APGs which assign shared roles to all players and let them control the game together through channel voters during the gaming process, DPGs assign an independent identity to each player and allow them to play games independently through inputting respective control commands in the live streaming channels. These innovations in player identity and control allow DPG players to interact freely with each other and also reduce the players' reliance on streamers. In addition, the importance of dynamics between streamers and players also changed, which is no longer a design focus in DPG channels. Unlike most APGs with inefficient controllers in previous studies, the streamers no longer need to stay in the DPG channels with players. Instead, the players can freely interact with each other and enjoy themselves, which is more similar to regular online games. Because of the series of design features aforementioned, most of the DPGs are no longer adapted from traditional single-player games (e.g., twitch Pokémon) but original games with innovative interactions between players. \textbf{In summary, DPGs are special APGs that allow channel players to freely control their characters and interact with each other through inputting comments in the channels. The players could be assigned independent identities and directly enjoy themselves in DPG channels instead of relying on streamers to provide feedback or voters to push the games forward.}

\subsection{Special Interactions Among Players Afforded by DPG Channels}

During the interview process, we found that DPGs could offer several special interactions among players. On the one hand, the independent identity assigned to players in the channels allows them to freely interact with each other, which enables special connections to be built among different groups of players (e.g., supervision from skilled players towards novice players). On the other hand, the low reliance on steamers forces the players to spontaneously manage the channels and maintain the order of channels.

\subsubsection{Supervision from Skilled Players Towards Novice Players}

The independent identity assigned by public live streaming channels also causes novice DPG players to accept supervision from skilled players. Unlike regular games, all the input commands of DPG players are public in the channels. For beginners in DPG channels, it is common for them to be judged by other skilled players. Especially in group fighting DPG games, their teammates may blame them for input mistakes or failed corporations. The criticism messages also appear on the list of chat messages of the channel publicly, leading to more criticisms from other players, e.g., \emph{``In some DPG channels, it is common that unexpected critical comments pop up because of some novice players' input.''} (P1). This may be offensive and discouraging to some new players since they are still learning the games and it is hard for them to avoid mistakes, e.g., \emph{``It is hard for novice players to remember the rules immediately. The criticism may embarrass them .''} (P1). In addition, some players see the criticisms from skilled players as part of the player-to-player interactions and have successfully adapted to them. Such public criticisms do not shame these players but consider them as normal inter-player instructions, e.g., \emph{``It is just like an important part of the game. The novice players can realize their mistakes from others' criticism"} (P7). Skilled players often take on the role of commanders in the DPG channels, where they are more active in speaking up about their thoughts about their strategies for the present game. In addition, they are also more eager to correct the mistakes of novice players.  

\subsubsection{Tension Between Paying Players and Non-paying Players}

The mixture of independent characters and shared gaming live streaming channels also intensifies the competition among DPG players and their influence on each other. Virtual tool battling is one common phenomenon in DPG channels. Some players purchase virtual tools on the platform to get potent weapons to defend their opponents. Many interviewees are curious about what influence or buff these special tools can bring to the channels when they buy them for the first time, e.g., \emph{``I just want to know what will happen after I buy them.''}(P1). Once they are satisfied with the superior convenience of tools, they may be addicted to buying tools, e.g., \emph{``I remember once I spent 1,500 in one game. I found out that a player from the other camp had purchased many props, and I wanted to compete with them. ''} (P8). For players unwilling to spend money on virtual tools, the purchasable tools partly destroy the fairness of DPG because all players need to share the channel. 

Interestingly, many interviewees reported that they turn to channels where tools-buying players do not enjoy too many advantages, e.g., \emph{``I have used the free props, but they are not very useful. There are always people willing to pay for awesome props, which makes the whole game dependent on this group of paying players and makes the game less fun.''} (P4) and \emph{``It takes away the gaming experience from players who do not spend money.''} (P9). To safeguard the gaming experience of free players, some game developers design special protection mechanisms for them and help them beat paying players, such as providing free props to non-paying players and setting limits on the number of times players could use props. These mechanisms narrow the gap between paying players and non-paying players, e.g., \emph{``The developers have designed a protection mechanism to protect non-paying players. The non-paying players would be given extra props when they are about to lose.''} (P5). This finding reveals that players of different economic conditions 
are easily diverted as they may choose their own most suitable gaming environment (e.g., different channels in DPGs). In addition, this finding also reveals that the game designers could adjust the convenience brought by paid props to attract their target players (e.g., strong and expensive props to attract more paying players).

\subsubsection{Maintaining the Order of Channels Together}

The low reliance of DPG channels on streamers also raises the importance of the player to maintain the live streaming channels spontaneously. We found that the DPG players always need to safeguard their gaming experience through connection with developers outside the channels and collaboration with each other inside the channels. For example, P5 told us that the game sometimes fails to work because of bugs and that they must go to the community regularly to call the developers to fix the problem and allow the games to continue. Another daily challenge happening in the DPG channels is fighting unruly players. Unlike APG, each player in DPG can input effective instructions, which the systems would execute, e.g., 

\begin{quote}
    \emph{``I remember one player once made the channel full of soldiers by buying gifts. Then the developer's computer crashed suddenly and can not work anymore.''} (P9). 
\end{quote}

\noindent In order to maintain the order of channels, some players spontaneously gather together and make use of game rules to fight malicious players; they may deliberately creating obstacles to malicious players during game playing, e.g., \emph{``I am very proud of my involvement in punishing these malicious players ... When I find a cheater appear in the channel, I may make use of my game credits to embarrass the cheater inside the game.''} (P9).

\subsection{Unique Player Experience Brought by DPG Channels}

During the interview process, we found that some unique experience could be afforded by DPGs based on their unique features compared to crowd-based APGs in \autoref{comparison}, which could be a more realistic and immersive gaming experience. Low reliance on steamers also allows DPG channel owners to focus on optimizing the games for a better gaming experience instead of hosting the channels, further strengthening the players' identity towards the DPG channels. However, the independent identity assigned to players may worsen the game's latency problem and reduce the whole channel's stability, negatively influencing the player experience.

\subsubsection{Positive Experience of DPG Players}

In this section, we introduce the positive player experience brought by the design attributes aforementioned, which include a strong sense of engagement and achievement, close connection with developers, and the freedom to play games anytime.  

\setlength{\parindent}{0cm} \paragraph{Stronger Sense of Engagement}

Compared to crowd-based APGs, the independent player identity and low reliance on streamers of DPG channels allow players to enjoy themselves instead of waiting for feedback from controllers or streamers. Some of our interviewees compared their gaming experience in DPG channels with ones in crowd-based APG channels. They all agreed that DPGs help them gain a stronger sense of involvement during the playing process instead of feeling like just an audience during the gaming process. \emph{Let's Genshin}, a crowd-based APG on Bilibili that allows players to play \emph{Genshin Impact} together, is mentioned by our interviewees, e.g.,

\begin{quote}

``\emph{Too many people are entering commands simultaneously in \emph{GenShin} channel, the delay is horrible...I think it is cooler to have one character per person, like one character per player fighting each other, which can lead to a more realistic gaming experience... I don't think I am playing a game in \emph{GenShin} channel.}'' (P6).

\end{quote}

\setlength{\parindent}{0cm} \paragraph{Connecting with Developers.}

The low reliance on streamers of DPG channels allows independent game developers to run their channels without having to show up to amuse players as regular live streaming streamers. Instead, they can focus on improving their games to provide a better gaming experience to players based on the feedback collected from players, which establishes a special connection between game developers and players. The content of most DPGs is not stable. Instead, they tend to change from day to day. It is common for DPG developers to keep adding new elements to the DPG (e.g. new characters, new commands, or new Auxiliary props), especially for new channels (e.g., \emph{Army Fight}) since the developers could receive fresh feedback from player communities (P5). Then the developers could optimize the games iteratively based on these feedback. Some DPGs may be homogeneous initially but gradually evolve in different directions, increasing the diversity of DPG channels. For example, some DPG channels started as marble fight games. However, they gradually developed into diverse styles, including cartoon-style marble fights, football-style marble fights, war-style marble fights, etc. Compared to APGs, where the content is primarily copied from traditional games, DPGs allow players to identify with developers who are constantly making improvements to the game and thus more willing to stay in the channels, e.g.,

\begin{quote}
\emph{``We identify with the developer because he can show us that he is working hard to make a new game. That is what identifies a developer ... That is why I am willing to stay in this DPG channel, and I am willing to help the developers as much as I can.''} (P8). 
\end{quote}

\setlength{\parindent}{0cm} \paragraph{Anytime for Fun with Low Burden}

The low reliance on streamers also allows DPG channels to be online 24 hours a day, which makes it possible for players to have fun at any time instead of following the schedules of streamers. When playing regular games, players must keep controlling their mouse, keyboard, or controller and concentrate on the game playing during the whole game playing process. However, in DPG, the cognitive load of game playing becomes relatively low, e.g., ``\textit{I only need an emoji to control the character to perform a series of actions set by the DPG creators instead of typing on the keyboards repeatedly}'' (P1). For some fighting DPG, one round of the game may only last for several minutes. This also reduces the burden on players. They can pick up their phones and start a game in their free time. This is very attractive for players with a busy schedule or lack of entertainment options, e.g., \emph{``I was bored and had nothing to do after work but stay inside my dorm until I saw such mixture of live streaming and gaming.''} (P3).

\subsubsection{Negative Experience of DPG Players}

In this section, we introduce the negative DPG player experience, which mainly includes channel latency and instability.

\setlength{\parindent}{0cm} 

\paragraph{Higher Latency and Instability}
Consistent with prior studies in APG \cite{glickman2018design}, players in DPG channels also agree that the latency problem negatively impacts their gaming experience. The independent identities assigned to players also deteriorated channels' latency and stability issues. Because all the players in the DPG channels could create characters and may enter a large number of commands at the same time, which brings a huge burden to the channel owner's device (e.g., the device that receives the instructions from channels and executes the game, which is often the private computer of channel owners). Our interviewees highlight that as the number of players in the channel increases, the latency problem worsens. Some DPG channel owners limit the number of players involved in the game to reduce the latency. However, this also brings down the player experience, e.g.,

\begin{quote}

\emph{"When I entered a comment, I could not see my action immediately from the screen because the latency problem was so bad. In some games that can accommodate many people, I need to wait ten seconds to see my command take effect. The werewolf live room would limit the number of players, which made it often hard for me to sign up."} (P4).

\end{quote}

The high influx of players not only worsens the latency problems appearing in regular APG channels but also reduces the game stability and causes the channel owners' computers to be more likely to crash and makes the whole game completely unplayable under extreme circumstances, e.g., 

\begin{quote}

\emph{"The whole computer screen was full of players, and then the channel owner's computer suddenly crashed, which also caused that laptop to break down. Now the channel owner has to use other's computer to continue the game."} (P9).

\end{quote}

This is although some of our interviewees expressed a desire to have dedicated servers to execute the DPGs rather than just relying on the personal computers of the channel owners (P4). As of February 2023, Bilibili still needs to arrange dedicated gaming servers for the DPGs channels.

\section{DISCUSSION}

Our results provide a nuanced understanding of the practice and experience of DPG players. Our observation and findings complement prior research in audience participation games, highlighting game players’ independence for a better gaming experience. We now reflect on the inspirations from DPGs for the design of APGs and the impact of DPGs on game players susceptible to the burden brought by general gaming process. We further offer several design implications for DPG and APG channels for a better gaming experience.

\subsection{An Alternative for Players Sensitive to Gaming Burden}

Although prior work has discussed players' motivation in APG channels, most of them only focus on the novel gaming experience (e.g., sense of power \cite{seering2017audience}) brought by APGs rather than the characteristics of these players. Our work revealed that some DPG players might be susceptible to the burden of the regular gaming process (e.g., effort required for game controlling). Previous studies in gaming have discussed these marginal game players, who may rely on outdated gaming devices \cite{lomas2013power}, face busy daily schedules and be physically weak \cite{radhakrishnan2020personalizing}. These players may completely avoid games with high device and technical requirements \cite{hasan2010multiplayer}. Some scholars in HCI tried to help these marginal players by building low-end device friendly systems \cite{laikari2010gaming} or proposing novel game design \cite{lomas2013power}. Our findings revealed that DPGs are a great alternative to help them enjoy gaming while saving extra effort. On the other hand, the deeper gamification of the DPG channels compared to APGs has dramatically enhanced the player gaming experience. According to the interactivity spectrum theory \cite{striner2019spectrum}, DPG players enjoy a more advanced level of interaction since each of them can act as not only a "viewer" in traditional live streaming but a "performer" (e.g., independent character) rather than an "influencer" in APG (e.g., vote for instructions). The features inherited from live streaming also enable players to enjoy a unique gaming experience. Unlike regular multi-player games, all player actions and communication are completely transparent and public in channels. However, our work reveals that it also makes it easy for less skilled players to suffer crowd criticism, a crucial privacy issue that previous scholars in APG studies always ignore.

\subsection{Deeper Integration of Game within Live Streaming}

Our findings reveal that DPGs can provide a stronger sense of participation to players than APGs (P6), which is because of the independent control and identity assigned to players. According to the interactivity spectrum theory \cite{striner2019spectrum}, DPG players enjoy a more advanced level of interaction since each of them can act as a ``performer'' (e.g., control independent characters and interact with each other) rather than only an ``influnencer'' in APGs (e.g., vote for instructions). The features inherited from live streaming also enable players to enjoy a unique gaming experience. Unlike regular multi-player games, all player actions and communication are completely transparent and public in channels as online comments. Although this allows new players to catch up quickly with other players by studying and copying their commands, it also makes it easy for less skilled players to suffer from crowd criticism. Gaming researchers have discussed the importance of player privacy management in video games \cite{tally2021protect}. However, most of their work only focus on regular video games instead of extremely transparent gaming environment like DPGs. Our findings revealed that DPGs could be an entry point for research into player privacy in extreme conditions. Our work also revealed that the mixture of public gaming space and individual independent characters intensifies the competition among channel players. Previous studies in computer games have stated that players may experience more negative feelings when they compete against others \cite{lim2010computer}, which is consistent with our finding that some non-paying players may search for appropriate channels which could safeguard their gaming experience. In summary, DPG is a novel integration of live streaming and games, which deserves more in-depth thinking by game designers to improve the player experience in this unique context and offers opportunities to game researchers to conduct further research.

\subsection{Design Implications}

Based on our results and discussion, we now provide design implications for future APG and DPG design, which include rethinking the role of streamers, collaboration with live streaming websites, and player privacy management.    

\setlength{\parindent}{0cm} \subsubsection{Rethinking the Role of Streamers in the Design of APGs}
Prior researchers have presented six key challenges for the design of APGs \cite{glickman2018design}, which include latency, screen sharing dilemma, player attention management, agency, relationship and shifting schedules between players and streamers. These key challenges significantly impact player experience and deserve careful consideration by designers. We found that DPGs could provide a potential idea for designers to rethink the role of streamers during the game design process. For the ideal APG design in previous studies, streamers must always be present and play with channel audiences as opponents or leading roles \cite{seering2017audience,glickman2018design}. One of the reasons may be that the comments of audiences are always different \cite {lessel2017crowdchess}, which means the output of controllers aggregating the comments from players may not drive the games forward efficiently. Therefore, streamers must be part of the game to safeguard the channel player experience. However, the streamer/audience format leads to another screen-sharing challenge, which refers to the shared screen needing to provide separate information (e.g., different views in one game) to both the streamers and the audiences. In addition, the streamer/audience format also requires the designers to reconfigure the relationship between streamers and audiences during the gaming process since the typical streamer-viewer model may not work in APG channels \cite{lessel2017crowdchess}. Meanwhile, the different schedules for streamers and audiences may also negatively affect the gaming process since the designers must consider how the game continues if some audiences leave the channels during game playing \cite{lessel2017crowdchess}. Our work revealed that DPGs may provide a novel idea for designers when facing these design challenges, which directly removes the controller and transfers independent control to each player. These design innovations make the live streaming channels less dependent on the streamers and help designers avoid thinking about the impact of streamers on the player's gaming experience. To our surprise, the collected data reveals that the control distribution to each player avoids the low efficient controller issues in regular APG channels and allows players to have a complete online gaming experience, including a stronger sense of involvement and achievement. Getting rid of the controller also optimizes the agency issues in regular APG channels since the common controllers, such as the voting system, reduce the sense of agency for the individual player significantly when the number of channel players increases \cite{lessel2017crowdchess}. In addition, some of the DPG design innovations, such as mapping player chat in the channel to in-game communication, also bring new ideas to APG designers to rethink the problem of player attention splitting issues during game playing process since the players can only focus on the main screen to chat, interact and strategies with other players without having to split attention to the chat box or other interfaces. In summary, DPG brings new design philosophy to APG design and inspires designers to deliver a complete gaming experience to channel players.

\setlength{\parindent}{0cm} \subsubsection{Collaborating with Live Streaming Websites}

As depicted in 4.2.3, DPG channel owners can hardly take extremely effective measures against malicious players. One reason may be they don't have access to the live streaming servers. It is, therefore, necessary for Bilibili to provide more comprehensive support to DPG channel owners. For example, the website could ban players with a specific IP address from logging in to the channels via the list of malicious accounts submitted by the channel owners. In addition, the live streaming websites should also introduce smarter strategies to guarantee the players' experience (e.g., Detecting malicious players who enter a large number of invalid commands) instead of relying on the extra effort from DPG players. The website could also provide dedicated servers for executing the DPGs provided by channel owners instead of relying on the channel owners' personal computers, which could significantly improve the stability of DPG channels and the player experience. In summary, we believe DPG designers should collaborate with live streaming websites deeply and introduce dedicated servers to provide a better experience to DPG players.

\setlength{\parindent}{0cm} \subsubsection{Designing for Player Privacy Management in DPGs}

Because of the completely transparent playing environment of DPG channels, unskilled players are often vulnerable to being criticized by skilled players, as discussed in 4.2.1. Although some players indicated they could tolerate this, we still find some novice players are afraid of being criticized. It is, therefore, necessary to provide privacy options to allow players to decide whether they want to make their comments public and whether they want to join the game anonymously. We believe this simple feature could ensure players' privacy is not compromised without affecting their gaming experience. In addition, the DPG designers could also arrange mandatory privacy protection for novice players who recently joined the channels since they are most vulnerable to attacks from others. For example, the designers could make the inputs of novice players private in the channels while they are familiarizing themselves with the games (e.g., the First 10 minutes after joining the channels.). We believe these privacy management strategies could make the channels more friendly for novice players.

\subsection{Limitations and Future Work}

We now note several limitations of our study that should be considered when interpreting it. Firstly, the interviewees are skewed young and highly educated. Further research is needed to test our understanding of the DPG community with a more diverse sample. Due to the nature of this study, causal generalizability of the findings could not be made. Interviewing different stakeholders, such as malicious players and website staff responsible for supporting DPG, could reveal more insights about DPG.

\section{CONCLUSION}

In this paper, we report the player practices and experience afforded by DPGs, one special kind of APGs allowing players to play games independently in live streaming channels. We conducted observation towards 50 DPG channels and interviewed 13 DPG players. Our work revealed that DPGs are mainly distinct from APGs with collaborative control in control and assigned the identity of channel players, interaction among channel players, and feedback from channels provided to players. Innovations in these attributes allow DPG players to have a unique player experience, which mainly includes a stronger sense of engagement and achievement and builds unique connections between game developers and players. However, the privacy issue of DPGs may also bring a negative influence on player experience. We consider that these innovations may bring new ideas to APG designers to improve design space and rethink the design challenges proposed by previous researchers. On the other hand, our work also revealed that DPGs may be a proper choice for marginal game players. Our work suggests opportunities to provide a complete gaming experience to players who can hardly afford high-end devices and spare time for gaming. Taken together, we see DPGs as valuable to explore for providing players with a more complete and immersive experience through live streaming channels. We look forward to more explorations of designing DPGs, as well as to new game design ideas inspired by DPGs.

\section*{ACKNOWLEDGEMENTS}
We sincerely thank all the DPG players who participated in our interviews. We are grateful for their sincere and detailed sharing concerning their gaming experience.      

\bibliographystyle{ACM-Reference-Format}
\bibliography{sample-base}


\begin{thebibliography}{54}


\ifx \showCODEN    \undefined \def \showCODEN     #1{\unskip}     \fi
\ifx \showDOI      \undefined \def \showDOI       #1{#1}\fi
\ifx \showISBNx    \undefined \def \showISBNx     #1{\unskip}     \fi
\ifx \showISBNxiii \undefined \def \showISBNxiii  #1{\unskip}     \fi
\ifx \showISSN     \undefined \def \showISSN      #1{\unskip}     \fi
\ifx \showLCCN     \undefined \def \showLCCN      #1{\unskip}     \fi
\ifx \shownote     \undefined \def \shownote      #1{#1}          \fi
\ifx \showarticletitle \undefined \def \showarticletitle #1{#1}   \fi
\ifx \showURL      \undefined \def \showURL       {\relax}        \fi
\providecommand\bibfield[2]{#2}
\providecommand\bibinfo[2]{#2}
\providecommand\natexlab[1]{#1}
\providecommand\showeprint[2][]{arXiv:#2}

\bibitem[Arsenault(2009)]%
        {arsenault2009video}
\bibfield{author}{\bibinfo{person}{Dominic Arsenault}.}
  \bibinfo{year}{2009}\natexlab{}.
\newblock \showarticletitle{Video game genre, evolution and innovation}.
\newblock \bibinfo{journal}{\emph{Eludamos: Journal for computer game culture}}
  \bibinfo{volume}{3}, \bibinfo{number}{2} (\bibinfo{year}{2009}),
  \bibinfo{pages}{149--176}.
\newblock


\bibitem[Bean(2016)]%
        {Choice}
\bibfield{author}{\bibinfo{person}{Studio Bean}.}
  \bibinfo{year}{2016}\natexlab{}.
\newblock \showarticletitle{Choice Chamber.}
\newblock
\urldef\tempurl%
\url{http://choicechamber.com/sub}
\showURL{%
\tempurl}


\bibitem[Benford et~al\mbox{.}(2021)]%
        {benford2021producing}
\bibfield{author}{\bibinfo{person}{Steve Benford}, \bibinfo{person}{Paul
  Mansfield}, {and} \bibinfo{person}{Jocelyn Spence}.}
  \bibinfo{year}{2021}\natexlab{}.
\newblock \showarticletitle{Producing Liveness: The Trials of Moving Folk Clubs
  Online During the Global Pandemic}. In \bibinfo{booktitle}{\emph{Proceedings
  of the 2021 CHI Conference on Human Factors in Computing Systems}}.
  \bibinfo{pages}{1--16}.
\newblock


\bibitem[Braun and Clarke(2012)]%
        {braun2012thematic}
\bibfield{author}{\bibinfo{person}{Virginia Braun} {and}
  \bibinfo{person}{Victoria Clarke}.} \bibinfo{year}{2012}\natexlab{}.
\newblock \bibinfo{booktitle}{\emph{Thematic analysis.}}
\newblock \bibinfo{publisher}{American Psychological Association}.
\newblock


\bibitem[Cai et~al\mbox{.}(2018)]%
        {cai2018utilitarian}
\bibfield{author}{\bibinfo{person}{Jie Cai}, \bibinfo{person}{Donghee~Yvette
  Wohn}, \bibinfo{person}{Ankit Mittal}, {and} \bibinfo{person}{Dhanush
  Sureshbabu}.} \bibinfo{year}{2018}\natexlab{}.
\newblock \showarticletitle{Utilitarian and hedonic motivations for live
  streaming shopping}. In \bibinfo{booktitle}{\emph{Proceedings of the 2018 ACM
  international conference on interactive experiences for TV and online
  video}}. \bibinfo{pages}{81--88}.
\newblock


\bibitem[Cerratto-Pargman et~al\mbox{.}(2014)]%
        {cerratto2014understanding}
\bibfield{author}{\bibinfo{person}{Teresa Cerratto-Pargman},
  \bibinfo{person}{Chiara Rossitto}, {and} \bibinfo{person}{Louise Barkhuus}.}
  \bibinfo{year}{2014}\natexlab{}.
\newblock \showarticletitle{Understanding audience participation in an
  interactive theater performance}. In \bibinfo{booktitle}{\emph{Proceedings of
  the 8th Nordic Conference on Human-Computer Interaction: Fun, Fast,
  Foundational}}. \bibinfo{pages}{608--617}.
\newblock


\bibitem[Chen et~al\mbox{.}(2021)]%
        {chen2021towards}
\bibfield{author}{\bibinfo{person}{Yan Chen}, \bibinfo{person}{Walter~S
  Lasecki}, {and} \bibinfo{person}{Tao Dong}.} \bibinfo{year}{2021}\natexlab{}.
\newblock \showarticletitle{Towards supporting programming education at scale
  via live streaming}.
\newblock \bibinfo{journal}{\emph{Proceedings of the ACM on Human-Computer
  Interaction}} \bibinfo{volume}{4}, \bibinfo{number}{CSCW3}
  (\bibinfo{year}{2021}), \bibinfo{pages}{1--19}.
\newblock


\bibitem[Ducheneaut and Moore(2004)]%
        {ducheneaut2004social}
\bibfield{author}{\bibinfo{person}{Nicolas Ducheneaut} {and}
  \bibinfo{person}{Robert~J Moore}.} \bibinfo{year}{2004}\natexlab{}.
\newblock \showarticletitle{The social side of gaming: a study of interaction
  patterns in a massively multiplayer online game}. In
  \bibinfo{booktitle}{\emph{Proceedings of the 2004 ACM conference on Computer
  supported cooperative work}}. \bibinfo{pages}{360--369}.
\newblock


\bibitem[Emmerich et~al\mbox{.}(2021)]%
        {emmerich2021streaming}
\bibfield{author}{\bibinfo{person}{Katharina Emmerich}, \bibinfo{person}{Andrey
  Krekhov}, \bibinfo{person}{Sebastian Cmentowski}, {and} \bibinfo{person}{Jens
  Krueger}.} \bibinfo{year}{2021}\natexlab{}.
\newblock \showarticletitle{Streaming vr games to the broad audience: A
  comparison of the first-person and third-person perspectives}. In
  \bibinfo{booktitle}{\emph{Proceedings of the 2021 CHI Conference on Human
  Factors in Computing Systems}}. \bibinfo{pages}{1--14}.
\newblock


\bibitem[Faklaris et~al\mbox{.}(2016)]%
        {faklaris2016legal}
\bibfield{author}{\bibinfo{person}{Cori Faklaris}, \bibinfo{person}{Francesco
  Cafaro}, \bibinfo{person}{Sara~Anne Hook}, \bibinfo{person}{Asa Blevins},
  \bibinfo{person}{Matt O'Haver}, {and} \bibinfo{person}{Neha Singhal}.}
  \bibinfo{year}{2016}\natexlab{}.
\newblock \showarticletitle{Legal and ethical implications of mobile
  live-streaming video apps}. In \bibinfo{booktitle}{\emph{Proceedings of the
  18th International Conference on Human-Computer Interaction with Mobile
  Devices and Services Adjunct}}. \bibinfo{pages}{722--729}.
\newblock


\bibitem[Glickman et~al\mbox{.}(2018)]%
        {glickman2018design}
\bibfield{author}{\bibinfo{person}{Seth Glickman}, \bibinfo{person}{Nathan
  McKenzie}, \bibinfo{person}{Joseph Seering}, \bibinfo{person}{Rachel
  Moeller}, {and} \bibinfo{person}{Jessica Hammer}.}
  \bibinfo{year}{2018}\natexlab{}.
\newblock \showarticletitle{Design challenges for livestreamed audience
  participation games}. In \bibinfo{booktitle}{\emph{Proceedings of the 2018
  Annual Symposium on Computer-Human Interaction in Play}}.
  \bibinfo{pages}{187--199}.
\newblock


\bibitem[Guo and Fussell(2022)]%
        {guo2022s}
\bibfield{author}{\bibinfo{person}{Jiajing Guo} {and} \bibinfo{person}{Susan~R
  Fussell}.} \bibinfo{year}{2022}\natexlab{}.
\newblock \showarticletitle{" It's Great to Exercise Together on Zoom!":
  Understanding the Practices and Challenges of Live Stream Group Fitness
  Classes}.
\newblock \bibinfo{journal}{\emph{Proceedings of the ACM on Human-Computer
  Interaction}} \bibinfo{volume}{6}, \bibinfo{number}{CSCW1}
  (\bibinfo{year}{2022}), \bibinfo{pages}{1--28}.
\newblock


\bibitem[Hamilton et~al\mbox{.}(2014)]%
        {hamilton2014streaming}
\bibfield{author}{\bibinfo{person}{William~A Hamilton}, \bibinfo{person}{Oliver
  Garretson}, {and} \bibinfo{person}{Andruid Kerne}.}
  \bibinfo{year}{2014}\natexlab{}.
\newblock \showarticletitle{Streaming on twitch: fostering participatory
  communities of play within live mixed media}. In
  \bibinfo{booktitle}{\emph{Proceedings of the SIGCHI conference on human
  factors in computing systems}}. \bibinfo{pages}{1315--1324}.
\newblock


\bibitem[Hammad et~al\mbox{.}(2021)]%
        {hammad2021towards}
\bibfield{author}{\bibinfo{person}{Noor Hammad}, \bibinfo{person}{Erik
  Harpstead}, {and} \bibinfo{person}{Jessica Hammer}.}
  \bibinfo{year}{2021}\natexlab{}.
\newblock \showarticletitle{Towards Examining The Effects of Live Streaming an
  Educational Game}. In \bibinfo{booktitle}{\emph{Extended Abstracts of the
  2021 CHI Conference on Human Factors in Computing Systems}}.
  \bibinfo{pages}{1--6}.
\newblock


\bibitem[Hasan(2010)]%
        {hasan2010multiplayer}
\bibfield{author}{\bibinfo{person}{Shiblee~Imtiaz Hasan}.}
  \bibinfo{year}{2010}\natexlab{}.
\newblock \showarticletitle{Multiplayer gaming for low-end mobile phones:
  Gaming between basic mobile phones, handheld devices and computer platforms}.
  In \bibinfo{booktitle}{\emph{2010 2nd International IEEE Consumer Electronics
  Society's Games Innovations Conference}}. IEEE, \bibinfo{pages}{1--5}.
\newblock


\bibitem[Hsu and Chen(2009)]%
        {hsu2009exploring}
\bibfield{author}{\bibinfo{person}{Chun-Cheng Hsu} {and} \bibinfo{person}{Elvis
  Chih-Hsien Chen}.} \bibinfo{year}{2009}\natexlab{}.
\newblock \showarticletitle{Exploring the elements and design criteria of
  massively-multiplayer online role-playing game (MMORPG) interfaces}. In
  \bibinfo{booktitle}{\emph{International Conference on Human-Computer
  Interaction}}. Springer, \bibinfo{pages}{325--334}.
\newblock


\bibitem[Kaytoue et~al\mbox{.}(2012)]%
        {kaytoue2012watch}
\bibfield{author}{\bibinfo{person}{Mehdi Kaytoue}, \bibinfo{person}{Arlei
  Silva}, \bibinfo{person}{Lo{\"\i}c Cerf}, \bibinfo{person}{Wagner Meira~Jr},
  {and} \bibinfo{person}{Chedy Ra{\"\i}ssi}.} \bibinfo{year}{2012}\natexlab{}.
\newblock \showarticletitle{Watch me playing, i am a professional: a first
  study on video game live streaming}. In \bibinfo{booktitle}{\emph{Proceedings
  of the 21st international conference on world wide web}}.
  \bibinfo{pages}{1181--1188}.
\newblock


\bibitem[Kitty(2015)]%
        {Legend}
\bibfield{author}{\bibinfo{person}{Robot~Loves Kitty}.}
  \bibinfo{year}{2015}\natexlab{}.
\newblock \showarticletitle{Legend of Dungeon: Masters.}
\newblock
\urldef\tempurl%
\url{https://robotloveskitty.com/}
\showURL{%
\tempurl}


\bibitem[Laikari et~al\mbox{.}(2010)]%
        {laikari2010gaming}
\bibfield{author}{\bibinfo{person}{Arto Laikari}, \bibinfo{person}{Jukka-Pekka
  Laulajainen}, \bibinfo{person}{Audrius Jurgelionis}, \bibinfo{person}{Philipp
  Fechteler}, {and} \bibinfo{person}{Francesco Bellotti}.}
  \bibinfo{year}{2010}\natexlab{}.
\newblock \showarticletitle{Gaming platform for running games on low-end
  devices}. In \bibinfo{booktitle}{\emph{User Centric Media: First
  International Conference, UCMedia 2009, Venice, Italy, December 9-11, 2009,
  Revised Selected Papers 1}}. Springer, \bibinfo{pages}{259--262}.
\newblock


\bibitem[Leavitt(2014)]%
        {Pokemon}
\bibfield{author}{\bibinfo{person}{Alex Leavitt}.}
  \bibinfo{year}{2014}\natexlab{}.
\newblock \showarticletitle{TwitchPlayedPokemon: An Analysis of the
  Experimental Interactive Phenomenon}.
\newblock
\urldef\tempurl%
\url{https://gdcvault.com/play/1021438}
\showURL{%
\tempurl}


\bibitem[Lessel et~al\mbox{.}(2018)]%
        {lessel2018viewers}
\bibfield{author}{\bibinfo{person}{Pascal Lessel}, \bibinfo{person}{Maximilian
  Altmeyer}, {and} \bibinfo{person}{Antonio Kr{\"u}ger}.}
  \bibinfo{year}{2018}\natexlab{}.
\newblock \showarticletitle{Viewers' perception of elements used in game
  live-streams}. In \bibinfo{booktitle}{\emph{Proceedings of the 22nd
  International Academic Mindtrek Conference}}. \bibinfo{pages}{59--68}.
\newblock


\bibitem[Lessel et~al\mbox{.}(2017a)]%
        {lessel2017crowdchess}
\bibfield{author}{\bibinfo{person}{Pascal Lessel}, \bibinfo{person}{Alexander
  Vielhauer}, {and} \bibinfo{person}{Antonio Kr{\"u}ger}.}
  \bibinfo{year}{2017}\natexlab{a}.
\newblock \showarticletitle{CrowdChess: A System to Investigate Shared Game
  Control in Live-Streams}. In \bibinfo{booktitle}{\emph{Proceedings of the
  Annual Symposium on Computer-Human Interaction in Play}}.
  \bibinfo{pages}{389--400}.
\newblock


\bibitem[Lessel et~al\mbox{.}(2017b)]%
        {lessel2017expanding}
\bibfield{author}{\bibinfo{person}{Pascal Lessel}, \bibinfo{person}{Alexander
  Vielhauer}, {and} \bibinfo{person}{Antonio Kr{\"u}ger}.}
  \bibinfo{year}{2017}\natexlab{b}.
\newblock \showarticletitle{Expanding video game live-streams with enhanced
  communication channels: A case study}. In
  \bibinfo{booktitle}{\emph{Proceedings of the 2017 CHI conference on human
  factors in computing systems}}. \bibinfo{pages}{1571--1576}.
\newblock


\bibitem[Lim and Reeves(2010)]%
        {lim2010computer}
\bibfield{author}{\bibinfo{person}{Sohye Lim} {and} \bibinfo{person}{Byron
  Reeves}.} \bibinfo{year}{2010}\natexlab{}.
\newblock \showarticletitle{Computer agents versus avatars: Responses to
  interactive game characters controlled by a computer or other player}.
\newblock \bibinfo{journal}{\emph{International Journal of Human-Computer
  Studies}} \bibinfo{volume}{68}, \bibinfo{number}{1-2} (\bibinfo{year}{2010}),
  \bibinfo{pages}{57--68}.
\newblock


\bibitem[Lomas et~al\mbox{.}(2013)]%
        {lomas2013power}
\bibfield{author}{\bibinfo{person}{Derek Lomas}, \bibinfo{person}{Anuj Kumar},
  \bibinfo{person}{Kishan Patel}, \bibinfo{person}{Dixie Ching},
  \bibinfo{person}{Meera Lakshmanan}, \bibinfo{person}{Matthew Kam}, {and}
  \bibinfo{person}{Jodi~L Forlizzi}.} \bibinfo{year}{2013}\natexlab{}.
\newblock \showarticletitle{The power of play: Design lessons for increasing
  the lifespan of outdated computers}. In \bibinfo{booktitle}{\emph{Proceedings
  of the SIGCHI Conference on Human Factors in Computing Systems}}.
  \bibinfo{pages}{2735--2744}.
\newblock


\bibitem[Lu(2019)]%
        {livestreaming2020lu}
\bibfield{author}{\bibinfo{person}{Zhicong Lu}.}
  \bibinfo{year}{2019}\natexlab{}.
\newblock \showarticletitle{Live Streaming in China for Sharing Knowledge and
  Promoting Intangible Cultural Heritage}.
\newblock \bibinfo{journal}{\emph{Interactions}} \bibinfo{volume}{27},
  \bibinfo{number}{1} (\bibinfo{date}{dec} \bibinfo{year}{2019}),
  \bibinfo{pages}{58–63}.
\newblock
\showISSN{1072-5520}
\urldef\tempurl%
\url{https://doi.org/10.1145/3373145}
\showDOI{\tempurl}


\bibitem[Lu et~al\mbox{.}(2019)]%
        {ICH2019lu}
\bibfield{author}{\bibinfo{person}{Zhicong Lu}, \bibinfo{person}{Michelle
  Annett}, \bibinfo{person}{Mingming Fan}, {and} \bibinfo{person}{Daniel
  Wigdor}.} \bibinfo{year}{2019}\natexlab{}.
\newblock \showarticletitle{"I Feel It is My Responsibility to Stream":
  Streaming and Engaging with Intangible Cultural Heritage through
  Livestreaming}. In \bibinfo{booktitle}{\emph{Proceedings of the 2019 CHI
  Conference on Human Factors in Computing Systems}} (Glasgow, Scotland Uk)
  \emph{(\bibinfo{series}{CHI '19})}. \bibinfo{publisher}{Association for
  Computing Machinery}, \bibinfo{address}{New York, NY, USA},
  \bibinfo{pages}{1–14}.
\newblock
\showISBNx{9781450359702}
\urldef\tempurl%
\url{https://doi.org/10.1145/3290605.3300459}
\showDOI{\tempurl}


\bibitem[Lu et~al\mbox{.}(2018a)]%
        {streamwiki2018lu}
\bibfield{author}{\bibinfo{person}{Zhicong Lu}, \bibinfo{person}{Seongkook
  Heo}, {and} \bibinfo{person}{Daniel~J. Wigdor}.}
  \bibinfo{year}{2018}\natexlab{a}.
\newblock \showarticletitle{StreamWiki: Enabling Viewers of Knowledge Sharing
  Live Streams to Collaboratively Generate Archival Documentation for Effective
  In-Stream and Post Hoc Learning}.
\newblock \bibinfo{journal}{\emph{Proc. ACM Hum.-Comput. Interact.}}
  \bibinfo{volume}{2}, \bibinfo{number}{CSCW}, Article \bibinfo{articleno}{112}
  (\bibinfo{date}{nov} \bibinfo{year}{2018}), \bibinfo{numpages}{26}~pages.
\newblock
\urldef\tempurl%
\url{https://doi.org/10.1145/3274381}
\showDOI{\tempurl}


\bibitem[Lu et~al\mbox{.}(2021a)]%
        {streamsketch2021lu}
\bibfield{author}{\bibinfo{person}{Zhicong Lu}, \bibinfo{person}{Rubaiat~Habib
  Kazi}, \bibinfo{person}{Li-yi Wei}, \bibinfo{person}{Mira Dontcheva}, {and}
  \bibinfo{person}{Karrie Karahalios}.} \bibinfo{year}{2021}\natexlab{a}.
\newblock \showarticletitle{StreamSketch: Exploring Multi-Modal Interactions in
  Creative Live Streams}.
\newblock \bibinfo{journal}{\emph{Proc. ACM Hum.-Comput. Interact.}}
  \bibinfo{volume}{5}, \bibinfo{number}{CSCW1}, Article \bibinfo{articleno}{58}
  (\bibinfo{date}{apr} \bibinfo{year}{2021}), \bibinfo{numpages}{26}~pages.
\newblock
\urldef\tempurl%
\url{https://doi.org/10.1145/3449132}
\showDOI{\tempurl}


\bibitem[Lu et~al\mbox{.}(2021b)]%
        {vtuber2021lu}
\bibfield{author}{\bibinfo{person}{Zhicong Lu}, \bibinfo{person}{Chenxinran
  Shen}, \bibinfo{person}{Jiannan Li}, \bibinfo{person}{Hong Shen}, {and}
  \bibinfo{person}{Daniel Wigdor}.} \bibinfo{year}{2021}\natexlab{b}.
\newblock \showarticletitle{More Kawaii than a Real-Person Live Streamer:
  Understanding How the Otaku Community Engages with and Perceives Virtual
  YouTubers}. In \bibinfo{booktitle}{\emph{Proceedings of the 2021 CHI
  Conference on Human Factors in Computing Systems}} (Yokohama, Japan)
  \emph{(\bibinfo{series}{CHI '21})}. \bibinfo{publisher}{Association for
  Computing Machinery}, \bibinfo{address}{New York, NY, USA}, Article
  \bibinfo{articleno}{137}, \bibinfo{numpages}{14}~pages.
\newblock
\showISBNx{9781450380966}
\urldef\tempurl%
\url{https://doi.org/10.1145/3411764.3445660}
\showDOI{\tempurl}


\bibitem[Lu et~al\mbox{.}(2018b)]%
        {lu2018you}
\bibfield{author}{\bibinfo{person}{Zhicong Lu}, \bibinfo{person}{Haijun Xia},
  \bibinfo{person}{Seongkook Heo}, {and} \bibinfo{person}{Daniel Wigdor}.}
  \bibinfo{year}{2018}\natexlab{b}.
\newblock \showarticletitle{You watch, you give, and you engage: a study of
  live streaming practices in China}. In \bibinfo{booktitle}{\emph{Proceedings
  of the 2018 CHI conference on human factors in computing systems}}.
  \bibinfo{pages}{1--13}.
\newblock


\bibitem[Lytle et~al\mbox{.}(2020)]%
        {lytle2020toward}
\bibfield{author}{\bibinfo{person}{Chance Lytle}, \bibinfo{person}{Parker
  Ramsey}, \bibinfo{person}{Joey Yeo}, \bibinfo{person}{Trace Dressen},
  \bibinfo{person}{Dong~hyun Kang}, \bibinfo{person}{Brenda~Bakker Harger},
  {and} \bibinfo{person}{Jessica Hammer}.} \bibinfo{year}{2020}\natexlab{}.
\newblock \showarticletitle{Toward live streamed improvisational game
  experiences}. In \bibinfo{booktitle}{\emph{Proceedings of the Annual
  Symposium on Computer-Human Interaction in Play}}. \bibinfo{pages}{148--159}.
\newblock


\bibitem[Moher et~al\mbox{.}(2009)]%
        {moher2009preferred}
\bibfield{author}{\bibinfo{person}{David Moher}, \bibinfo{person}{Alessandro
  Liberati}, \bibinfo{person}{Jennifer Tetzlaff}, \bibinfo{person}{Douglas~G
  Altman}, {and} \bibinfo{person}{PRISMA Group*}.}
  \bibinfo{year}{2009}\natexlab{}.
\newblock \showarticletitle{Preferred reporting items for systematic reviews
  and meta-analyses: the PRISMA statement}.
\newblock \bibinfo{journal}{\emph{Annals of internal medicine}}
  \bibinfo{volume}{151}, \bibinfo{number}{4} (\bibinfo{year}{2009}),
  \bibinfo{pages}{264--269}.
\newblock


\bibitem[Nguyen et~al\mbox{.}(2020)]%
        {9231771}
\bibfield{author}{\bibinfo{person}{Ngoc~Cuong Nguyen}, \bibinfo{person}{Ruck
  Thawonmas}, \bibinfo{person}{Pujana Paliyawan}, {and} \bibinfo{person}{Hai~V.
  Pham}.} \bibinfo{year}{2020}\natexlab{}.
\newblock \showarticletitle{JUSTIN: An Audience Participation Game With A
  Purpose for Collecting Descriptions for Artwork Images}. In
  \bibinfo{booktitle}{\emph{2020 IEEE Conference on Games (CoG)}}.
  \bibinfo{pages}{344--350}.
\newblock
\urldef\tempurl%
\url{https://doi.org/10.1109/CoG47356.2020.9231771}
\showDOI{\tempurl}


\bibitem[Paliyawan et~al\mbox{.}(2020)]%
        {paliyawan2020towards}
\bibfield{author}{\bibinfo{person}{Pujana Paliyawan}, \bibinfo{person}{Kingkarn
  Sookhanaphibarn}, \bibinfo{person}{Worawat Choensawat}, {and}
  \bibinfo{person}{Ruck Thawonmas}.} \bibinfo{year}{2020}\natexlab{}.
\newblock \showarticletitle{Towards Social Facilitation in Audience
  Participation Games: Fighting Game AIs whose Strength Depends on Audience
  Responses}. In \bibinfo{booktitle}{\emph{2020 IEEE Conference on Games
  (CoG)}}. IEEE, \bibinfo{pages}{686--689}.
\newblock


\bibitem[Paliyawan et~al\mbox{.}(2022)]%
        {paliyawan2022audience}
\bibfield{author}{\bibinfo{person}{Pujana Paliyawan}, \bibinfo{person}{Ruck
  Thawonmas}, \bibinfo{person}{Kingkarn Sookhanaphibarn}, {and}
  \bibinfo{person}{Worawat Choensawat}.} \bibinfo{year}{2022}\natexlab{}.
\newblock \showarticletitle{Audience Participation Fighting Game: making an APG
  using a concept of social facilitation}.
\newblock  (\bibinfo{year}{2022}).
\newblock


\bibitem[Radhakrishnan et~al\mbox{.}(2020)]%
        {radhakrishnan2020personalizing}
\bibfield{author}{\bibinfo{person}{Kavita Radhakrishnan},
  \bibinfo{person}{Thomas Baranowski}, \bibinfo{person}{Matthew O'Hair},
  \bibinfo{person}{Catherine~A Fournier}, \bibinfo{person}{Cathy~B Spranger},
  {and} \bibinfo{person}{Miyong~T Kim}.} \bibinfo{year}{2020}\natexlab{}.
\newblock \showarticletitle{Personalizing sensor-controlled digital gaming to
  self-management needs of older adults with heart failure: a qualitative
  study}.
\newblock \bibinfo{journal}{\emph{Games for Health Journal}}
  \bibinfo{volume}{9}, \bibinfo{number}{4} (\bibinfo{year}{2020}),
  \bibinfo{pages}{304--310}.
\newblock


\bibitem[Ramirez et~al\mbox{.}(2014)]%
        {ramirez2014twitch}
\bibfield{author}{\bibinfo{person}{Dennis Ramirez}, \bibinfo{person}{Jenny
  Saucerman}, {and} \bibinfo{person}{Jeremy Dietmeier}.}
  \bibinfo{year}{2014}\natexlab{}.
\newblock \showarticletitle{Twitch plays pokemon: a case study in big g games}.
  In \bibinfo{booktitle}{\emph{Proceedings of DiGRA}}. \bibinfo{pages}{3--6}.
\newblock


\bibitem[Saaty et~al\mbox{.}(2022)]%
        {saaty2022pokemon}
\bibfield{author}{\bibinfo{person}{Morva Saaty}, \bibinfo{person}{Derek Haqq},
  \bibinfo{person}{Mohammadreza Beyki}, \bibinfo{person}{Taha Hassan}, {and}
  \bibinfo{person}{D~SCOTT McCRICKARD}.} \bibinfo{year}{2022}\natexlab{}.
\newblock \showarticletitle{Pok{\'e}mon GO with Social Distancing: Social Media
  Analysis of Players' Experiences with Location-based Games}.
\newblock \bibinfo{journal}{\emph{Proceedings of the ACM on Human-Computer
  Interaction}} \bibinfo{volume}{6}, \bibinfo{number}{CHI PLAY}
  (\bibinfo{year}{2022}), \bibinfo{pages}{1--22}.
\newblock


\bibitem[Scheibe et~al\mbox{.}(2016)]%
        {scheibe2016information}
\bibfield{author}{\bibinfo{person}{Katrin Scheibe}, \bibinfo{person}{Kaja~J
  Fietkiewicz}, {and} \bibinfo{person}{Wolfgang~G Stock}.}
  \bibinfo{year}{2016}\natexlab{}.
\newblock \showarticletitle{Information behavior on social live streaming
  services}.
\newblock \bibinfo{journal}{\emph{Journal of Information Science Theory and
  Practice}} \bibinfo{volume}{4}, \bibinfo{number}{2} (\bibinfo{year}{2016}),
  \bibinfo{pages}{6--20}.
\newblock


\bibitem[Seering et~al\mbox{.}(2017)]%
        {seering2017audience}
\bibfield{author}{\bibinfo{person}{Joseph Seering}, \bibinfo{person}{Saiph
  Savage}, \bibinfo{person}{Michael Eagle}, \bibinfo{person}{Joshua Churchin},
  \bibinfo{person}{Rachel Moeller}, \bibinfo{person}{Jeffrey~P Bigham}, {and}
  \bibinfo{person}{Jessica Hammer}.} \bibinfo{year}{2017}\natexlab{}.
\newblock \showarticletitle{Audience participation games: Blurring the line
  between player and spectator}. In \bibinfo{booktitle}{\emph{Proceedings of
  the 2017 Conference on Designing Interactive Systems}}.
  \bibinfo{pages}{429--440}.
\newblock


\bibitem[Smith et~al\mbox{.}(2013)]%
        {smith2013live}
\bibfield{author}{\bibinfo{person}{Thomas Smith}, \bibinfo{person}{Marianna
  Obrist}, {and} \bibinfo{person}{Peter Wright}.}
  \bibinfo{year}{2013}\natexlab{}.
\newblock \showarticletitle{Live-streaming changes the (video) game}. In
  \bibinfo{booktitle}{\emph{Proceedings of the 11th european conference on
  Interactive TV and video}}. \bibinfo{pages}{131--138}.
\newblock


\bibitem[Striner et~al\mbox{.}(2019)]%
        {striner2019spectrum}
\bibfield{author}{\bibinfo{person}{Alina Striner}, \bibinfo{person}{Sasha
  Azad}, {and} \bibinfo{person}{Chris Martens}.}
  \bibinfo{year}{2019}\natexlab{}.
\newblock \showarticletitle{A spectrum of audience interactivity for
  entertainment domains}. In \bibinfo{booktitle}{\emph{International Conference
  on Interactive Digital Storytelling}}. Springer, \bibinfo{pages}{214--232}.
\newblock


\bibitem[Striner et~al\mbox{.}(2021)]%
        {striner2021mapping}
\bibfield{author}{\bibinfo{person}{Alina Striner}, \bibinfo{person}{Andrew~M
  Webb}, \bibinfo{person}{Jessica Hammer}, {and} \bibinfo{person}{Amy Cook}.}
  \bibinfo{year}{2021}\natexlab{}.
\newblock \showarticletitle{Mapping Design Spaces for Audience Participation in
  Game Live Streaming}. In \bibinfo{booktitle}{\emph{Proceedings of the 2021
  CHI Conference on Human Factors in Computing Systems}}.
  \bibinfo{pages}{1--15}.
\newblock


\bibitem[Tally et~al\mbox{.}(2021)]%
        {tally2021protect}
\bibfield{author}{\bibinfo{person}{Anne~Clara Tally}, \bibinfo{person}{Yu~Ra
  Kim}, \bibinfo{person}{Katreen Boustani}, {and} \bibinfo{person}{Christena
  Nippert-Eng}.} \bibinfo{year}{2021}\natexlab{}.
\newblock \showarticletitle{Protect and Project: Names, Privacy, and the
  Boundary Negotiations of Online Video Game Players}.
\newblock \bibinfo{journal}{\emph{Proceedings of the ACM on Human-Computer
  Interaction}} \bibinfo{volume}{5}, \bibinfo{number}{CSCW1}
  (\bibinfo{year}{2021}), \bibinfo{pages}{1--19}.
\newblock


\bibitem[Tang et~al\mbox{.}(2017)]%
        {tang2017perspectives}
\bibfield{author}{\bibinfo{person}{John~C Tang}, \bibinfo{person}{Funda
  Kivran-Swaine}, \bibinfo{person}{Kori Inkpen}, {and} \bibinfo{person}{Nancy
  Van~House}.} \bibinfo{year}{2017}\natexlab{}.
\newblock \showarticletitle{Perspectives on live streaming: Apps, users, and
  research}. In \bibinfo{booktitle}{\emph{Companion of the 2017 ACM Conference
  on Computer Supported Cooperative Work and Social Computing}}.
  \bibinfo{pages}{123--126}.
\newblock


\bibitem[Tang et~al\mbox{.}(2022)]%
        {ecommerce2022tang}
\bibfield{author}{\bibinfo{person}{Ningjing Tang}, \bibinfo{person}{Lei Tao},
  \bibinfo{person}{Bo Wen}, {and} \bibinfo{person}{Zhicong Lu}.}
  \bibinfo{year}{2022}\natexlab{}.
\newblock \showarticletitle{Dare to Dream, Dare to Livestream: How E-Commerce
  Livestreaming Empowers Chinese Rural Women}. In
  \bibinfo{booktitle}{\emph{Proceedings of the 2022 CHI Conference on Human
  Factors in Computing Systems}} (New Orleans, LA, USA)
  \emph{(\bibinfo{series}{CHI '22})}. \bibinfo{publisher}{Association for
  Computing Machinery}, \bibinfo{address}{New York, NY, USA}, Article
  \bibinfo{articleno}{297}, \bibinfo{numpages}{13}~pages.
\newblock
\showISBNx{9781450391573}
\urldef\tempurl%
\url{https://doi.org/10.1145/3491102.3517634}
\showDOI{\tempurl}


\bibitem[Wolff and Shen(2022)]%
        {wolff2022audience}
\bibfield{author}{\bibinfo{person}{Grace~H Wolff} {and} \bibinfo{person}{Cuihua
  Shen}.} \bibinfo{year}{2022}\natexlab{}.
\newblock \showarticletitle{Audience size, moderator activity, gender, and
  content diversity: Exploring user participation and financial commitment on
  Twitch. tv}.
\newblock \bibinfo{journal}{\emph{New Media \& Society}}
  (\bibinfo{year}{2022}), \bibinfo{pages}{14614448211069996}.
\newblock


\bibitem[Wu et~al\mbox{.}(2023)]%
        {ecommerce2023wu}
\bibfield{author}{\bibinfo{person}{Qunfang Wu}, \bibinfo{person}{Yisi Sang},
  \bibinfo{person}{Dakuo Wang}, {and} \bibinfo{person}{Zhicong Lu}.}
  \bibinfo{year}{2023}\natexlab{}.
\newblock \showarticletitle{Malicious Selling Strategies in Livestream
  E-Commerce: A Case Study of Alibaba’s Taobao and ByteDance’s TikTok}.
\newblock \bibinfo{journal}{\emph{ACM Trans. Comput.-Hum. Interact.}}
  \bibinfo{volume}{30}, \bibinfo{number}{3}, Article \bibinfo{articleno}{35}
  (\bibinfo{date}{jun} \bibinfo{year}{2023}), \bibinfo{numpages}{29}~pages.
\newblock
\showISSN{1073-0516}
\urldef\tempurl%
\url{https://doi.org/10.1145/3577199}
\showDOI{\tempurl}


\bibitem[Wu et~al\mbox{.}(2018)]%
        {wu2018danmaku}
\bibfield{author}{\bibinfo{person}{Qunfang Wu}, \bibinfo{person}{Yisi Sang},
  \bibinfo{person}{Shan Zhang}, {and} \bibinfo{person}{Yun Huang}.}
  \bibinfo{year}{2018}\natexlab{}.
\newblock \showarticletitle{Danmaku vs. forum comments: understanding user
  participation and knowledge sharing in online videos}. In
  \bibinfo{booktitle}{\emph{Proceedings of the 2018 ACM International
  Conference on Supporting Group Work}}. \bibinfo{pages}{209--218}.
\newblock


\bibitem[Wulf et~al\mbox{.}(2020)]%
        {wulf2020watching}
\bibfield{author}{\bibinfo{person}{Tim Wulf}, \bibinfo{person}{Frank~M
  Schneider}, {and} \bibinfo{person}{Stefan Beckert}.}
  \bibinfo{year}{2020}\natexlab{}.
\newblock \showarticletitle{Watching players: An exploration of media enjoyment
  on Twitch}.
\newblock \bibinfo{journal}{\emph{Games and culture}} \bibinfo{volume}{15},
  \bibinfo{number}{3} (\bibinfo{year}{2020}), \bibinfo{pages}{328--346}.
\newblock


\bibitem[Xiao et~al\mbox{.}(2020)]%
        {xiao2020random}
\bibfield{author}{\bibinfo{person}{Sijia Xiao}, \bibinfo{person}{Dana{\"e}
  Metaxa}, \bibinfo{person}{Joon~Sung Park}, \bibinfo{person}{Karrie
  Karahalios}, {and} \bibinfo{person}{Niloufar Salehi}.}
  \bibinfo{year}{2020}\natexlab{}.
\newblock \showarticletitle{Random, messy, funny, raw: Finstas as intimate
  reconfigurations of social media}. In \bibinfo{booktitle}{\emph{Proceedings
  of the 2020 CHI conference on human factors in computing systems}}.
  \bibinfo{pages}{1--13}.
\newblock


\bibitem[Yen et~al\mbox{.}(2023)]%
        {Yen2023Storychat}
\bibfield{author}{\bibinfo{person}{Ryan Yen}, \bibinfo{person}{Li Feng},
  \bibinfo{person}{Brinda Mehra}, \bibinfo{person}{Ching~Christie Pang},
  \bibinfo{person}{Siying Hu}, {and} \bibinfo{person}{Zhicong Lu}.}
  \bibinfo{year}{2023}\natexlab{}.
\newblock \showarticletitle{StoryChat: Designing a Narrative-Based Viewer
  Participation Tool for Live Streaming Chatrooms}. In
  \bibinfo{booktitle}{\emph{Proceedings of the 2023 CHI Conference on Human
  Factors in Computing Systems}} (Hamburg, Germany) \emph{(\bibinfo{series}{CHI
  '23})}. \bibinfo{publisher}{Association for Computing Machinery},
  \bibinfo{address}{New York, NY, USA}, Article \bibinfo{articleno}{795},
  \bibinfo{numpages}{18}~pages.
\newblock
\showISBNx{9781450394215}
\urldef\tempurl%
\url{https://doi.org/10.1145/3544548.3580912}
\showDOI{\tempurl}


\bibitem[Zhuang et~al\mbox{.}(2007)]%
        {zhuang2007player}
\bibfield{author}{\bibinfo{person}{Xinyu Zhuang}, \bibinfo{person}{Ashwin
  Bharambe}, \bibinfo{person}{Jeffrey Pang}, {and} \bibinfo{person}{Srinivasan
  Seshan}.} \bibinfo{year}{2007}\natexlab{}.
\newblock \showarticletitle{Player dynamics in massively multiplayer online
  games}.
\newblock \bibinfo{journal}{\emph{School of Computer Science, Carnegie Mellon
  University, Pittsburgh, Tech. Rep. CMU-CS-07-158}} (\bibinfo{year}{2007}),
  \bibinfo{pages}{30}.
\newblock


\end{thebibliography}

\end{document}